\documentclass[11pt]{article}

\usepackage[margin=1.1in]{geometry}
\usepackage{amsmath,amssymb,amsthm}
\usepackage{booktabs}
\usepackage{algorithm}
\usepackage{algpseudocode}
\usepackage{pgfplots}
\pgfplotsset{compat=1.18}
\pgfplotsset{
  discard if not/.style 2 args={
    x filter/.append code={
      \edef\tempa{\thisrow{#1}}
      \edef\tempb{#2}
      \ifx\tempa\tempb\else\fi
    }
  }
}
\usepackage[hidelinks]{hyperref}
\hypersetup{pdftitle={A Simple Randomized Estimator for Harmonic Numbers and Logarithms},
  pdfauthor={Ali Dasdan}}
\usepackage{placeins}
\gdef\eOneRows{%
$4^{1}$ & 2.0833 & 2.0832 & 0.0260 & 0.0257 & 1.3861 & 1.3863 & 0.01 & 0.948 & 0.950 \\
$4^{2}$ & 3.3807 & 3.3822 & 0.0442 & 0.0424 & 2.7741 & 2.7726 & 0.05 & 0.942 & 0.942 \\
$4^{3}$ & 4.7439 & 4.7459 & 0.0543 & 0.0558 & 4.1609 & 4.1589 & 0.05 & 0.952 & 0.952 \\
$4^{4}$ & 6.1243 & 6.1206 & 0.0682 & 0.0670 & 5.5414 & 5.5452 & 0.07 & 0.938 & 0.938 \\
$4^{5}$ & 7.5092 & 7.5041 & 0.0778 & 0.0766 & 6.9264 & 6.9315 & 0.07 & 0.952 & 0.952 \\
$4^{6}$ & 8.8951 & 8.8965 & 0.0864 & 0.0851 & 8.3192 & 8.3178 & 0.02 & 0.942 & 0.942 \\
$4^{7}$ & 10.2813 & 10.2844 & 0.0940 & 0.0929 & 9.7072 & 9.7041 & 0.03 & 0.946 & 0.946 \\
$4^{8}$ & 11.6676 & 11.6679 & 0.1015 & 0.1001 & 11.0907 & 11.0904 & 0.00 & 0.954 & 0.954 \\
}

\gdef\eFourRows{%
3 & 2 & 0.4055 & 0.4046 & 0.0154 & 0.0149 & 0.942 & 0.0273 & 0.0269 & 0.930 & $-3.7\times 10^{-4}$ \\
10 & 3 & 1.2040 & 1.2008 & 0.0297 & 0.0301 & 0.952 & 0.0435 & 0.0430 & 0.946 & $-9.7\times 10^{-5}$ \\
100 & 50 & 0.6931 & 0.6936 & 0.0263 & 0.0260 & 0.936 & 0.0802 & 0.0802 & 0.938 & $-1.2\times 10^{-9}$ \\
1000 & 1 & 6.9078 & 6.8999 & 0.0737 & 0.0764 & 0.952 & 0.0781 & 0.0764 & 0.944 & $-6.1\times 10^{-3}$ \\
1000 & 999 & 0.0010 & 0.0010 & 0.0010 & 0.0010 & 0.626 & 0.1057 & 0.1081 & 0.954 & $-3.3\times 10^{-17}$ \\
1001 & 1000 & 0.0010 & 0.0010 & 0.0010 & 0.0010 & 0.646 & 0.1107 & 0.1081 & 0.942 & $-3.3\times 10^{-17}$ \\
}

\gdef\eFiveRows{%
$4^{1}$ & 0.660 & 0.644 & 0.118 & 0.118 & 1.24 & 0.23 \\
$4^{2}$ & 1.796 & 1.829 & 0.606 & 0.625 & 1.24 & 0.42 \\
$4^{4}$ & 4.483 & 4.504 & 1.921 & 1.977 & 1.11 & 0.49 \\
$4^{6}$ & 7.250 & 7.030 & 3.303 & 3.225 & 1.09 & 0.50 \\
$4^{8}$ & 10.023 & 10.220 & 4.689 & 4.830 & 1.09 & 0.52 \\
}

\gdef\eSixBRows{%
1 & 0 & 1 & 0.0000 & 0.00000 & 0.00000 & 0.000 & 1.000 & 1.000 & $0$ \\
2 & 1 & 2 & 0.0000 & 0.00000 & 0.00000 & 0.000 & 1.000 & 1.000 & $0$ \\
3 & 1 & 2 & 0.4055 & 0.01530 & 0.01491 & 1.333 & 0.956 & 0.000 & $-3.7\times 10^{-4}$ \\
5 & 2 & 4 & 0.2231 & 0.01294 & 0.01265 & 1.200 & 0.936 & 0.000 & $-1.9\times 10^{-5}$ \\
7 & 2 & 4 & 0.5596 & 0.02077 & 0.02053 & 1.510 & 0.940 & 0.000 & $-2.8\times 10^{-5}$ \\
1024 & 10 & 1024 & 0.0000 & 0.00000 & 0.00000 & 0.000 & 1.000 & 1.000 & $0$ \\
1025 & 10 & 1024 & 0.0010 & 0.00097 & 0.00099 & 1.001 & 0.624 & 0.376 & $-3.0\times 10^{-17}$ \\
2047 & 10 & 1024 & 0.6927 & 0.02741 & 0.02630 & 1.692 & 0.940 & 0.000 & $-7.1\times 10^{-15}$ \\
2048 & 11 & 2048 & 0.0000 & 0.00000 & 0.00000 & 0.000 & 1.000 & 1.000 & $0$ \\
}

\gdef\eSixDRows{%
3 & 1 & 3 & 2 & $-3.7\times 10^{-4}$ & $5.2\times 10^{-4}$ & 0.2222 & 0.4055 & 1.3333 \\
3 & 2 & 6 & 4 & $-2.5\times 10^{-5}$ & $3.3\times 10^{-5}$ & 0.2989 & 0.4055 & 1.3667 \\
3 & 4 & 12 & 8 & $-1.6\times 10^{-6}$ & $2.0\times 10^{-6}$ & 0.3478 & 0.4055 & 1.3854 \\
3 & 16 & 48 & 32 & $-6.4\times 10^{-9}$ & $7.9\times 10^{-9}$ & 0.3902 & 0.4055 & 1.4003 \\
5 & 1 & 5 & 4 & $-1.9\times 10^{-5}$ & $3.3\times 10^{-5}$ & 0.1600 & 0.2231 & 1.2000 \\
5 & 2 & 10 & 8 & $-1.2\times 10^{-6}$ & $2.0\times 10^{-6}$ & 0.1888 & 0.2231 & 1.2111 \\
5 & 4 & 20 & 16 & $-7.5\times 10^{-8}$ & $1.3\times 10^{-7}$ & 0.2052 & 0.2231 & 1.2170 \\
5 & 16 & 80 & 64 & $-2.9\times 10^{-10}$ & $5.0\times 10^{-10}$ & 0.2185 & 0.2231 & 1.2216 \\
7 & 1 & 7 & 4 & $-2.8\times 10^{-5}$ & $3.3\times 10^{-5}$ & 0.4213 & 0.5596 & 1.5095 \\
7 & 2 & 14 & 8 & $-1.8\times 10^{-6}$ & $2.0\times 10^{-6}$ & 0.4851 & 0.5596 & 1.5337 \\
7 & 4 & 28 & 16 & $-1.1\times 10^{-7}$ & $1.3\times 10^{-7}$ & 0.5209 & 0.5596 & 1.5464 \\
7 & 16 & 112 & 64 & $-4.4\times 10^{-10}$ & $5.0\times 10^{-10}$ & 0.5497 & 0.5596 & 1.5563 \\
}

\gdef\eSevenARows{%
$3 \cdot 2^{1}$ & 0.2000 & $5.5\times 10^{-3}$ & $1.3\times 10^{-4}$ & $1.2\times 10^{-7}$ & $1.6\times 10^{-13}$ & $4.3\times 10^{-17}$ & $5.4\times 10^{-25}$ \\
$3 \cdot 2^{3}$ & 0.2000 & $5.5\times 10^{-3}$ & $1.3\times 10^{-4}$ & $1.2\times 10^{-7}$ & $1.6\times 10^{-13}$ & $3.1\times 10^{-16}$ & $5.4\times 10^{-25}$ \\
$3 \cdot 2^{7}$ & 0.2000 & $5.5\times 10^{-3}$ & $1.3\times 10^{-4}$ & $1.2\times 10^{-7}$ & $1.6\times 10^{-13}$ & $3.9\times 10^{-17}$ & $5.4\times 10^{-25}$ \\
$3 \cdot 2^{11}$ & 0.2000 & $5.5\times 10^{-3}$ & $1.3\times 10^{-4}$ & $1.2\times 10^{-7}$ & $1.6\times 10^{-13}$ & $1.4\times 10^{-15}$ & $5.4\times 10^{-25}$ \\
$3 \cdot 2^{15}$ & 0.2000 & $5.5\times 10^{-3}$ & $1.3\times 10^{-4}$ & $1.2\times 10^{-7}$ & $1.6\times 10^{-13}$ & $1.0\times 10^{-15}$ & $5.4\times 10^{-25}$ \\
1 & 0.0000 & $0$ & $0$ & $0$ & $0$ & $0$ & $0$ \\
$2^{10}$ & 0.0000 & $9.9\times 10^{-18}$ & $9.9\times 10^{-18}$ & $9.9\times 10^{-18}$ & $9.9\times 10^{-18}$ & $9.9\times 10^{-18}$ & $0$ \\
$2^{10}+1$ & $4.9\times 10^{-4}$ & $7.7\times 10^{-11}$ & $4.0\times 10^{-16}$ & $4.0\times 10^{-16}$ & $4.0\times 10^{-16}$ & $4.0\times 10^{-16}$ & $3.2\times 10^{-111}$ \\
$2^{11}-1$ & 0.3331 & $2.6\times 10^{-2}$ & $1.8\times 10^{-3}$ & $1.2\times 10^{-5}$ & $1.0\times 10^{-9}$ & $2.1\times 10^{-16}$ & $1.2\times 10^{-17}$ \\
$2^{11}$ & 0.0000 & $7.8\times 10^{-17}$ & $7.8\times 10^{-17}$ & $7.8\times 10^{-17}$ & $7.8\times 10^{-17}$ & $7.8\times 10^{-17}$ & $0$ \\
$2^{32}+1$ & $1.2\times 10^{-10}$ & $7.4\times 10^{-16}$ & $7.4\times 10^{-16}$ & $7.4\times 10^{-16}$ & $7.4\times 10^{-16}$ & $7.4\times 10^{-16}$ & $9.1\times 10^{-330}$ \\
$2^{53}-1$ & 0.3333 & $2.6\times 10^{-2}$ & $1.8\times 10^{-3}$ & $1.2\times 10^{-5}$ & $1.0\times 10^{-9}$ & $5.6\times 10^{-16}$ & $1.2\times 10^{-17}$ \\
$2^{63}-1$ & 0.3333 & $2.6\times 10^{-2}$ & $1.8\times 10^{-3}$ & $1.2\times 10^{-5}$ & $1.0\times 10^{-9}$ & $3.8\times 10^{-15}$ & $1.2\times 10^{-17}$ \\
}

\newtheorem{proposition}{Proposition}
\newtheorem{lemma}{Lemma}
\theoremstyle{remark}
\newtheorem{remark}{Remark}

\newcommand{\E}{\mathbb{E}}
\newcommand{\Var}{\operatorname{Var}}
\newcommand{\Cov}{\operatorname{Cov}}
\newcommand{\Hbar}{\bar{H}}
\newcommand{\Lhat}{\hat{L}}
\newcommand{\Lred}{\tilde{L}}
\title{A Simple Randomized Estimator for\\ Harmonic Numbers and Logarithms}
\author{
  Ali Dasdan \\
  KD Consulting \\
  Saratoga, CA, USA \\
  \texttt{alidasdan@gmail.com}
}
\date{Third version, October 2026. Versions 1 and 2 April 2017 (arXiv:1704.04538).}

\begin{document}
\maketitle

\begin{abstract}
Scanning a list of $x$ distinct numbers in uniformly random order for
its maximum replaces the running maximum $H_x$ times on average, where
$H_x$ is the $x$th harmonic number. Run backwards, this gives an
estimator: the replacement count of a random list is an unbiased Monte
Carlo estimate of $H_x$, and since $H_x = \ln x + \gamma + O(1/x)$, a
corrected estimate of $\ln x$. We give the exact variance of the
count, an error decomposition with sample-size rules, and an exact
sampler of the count using $H_x$ integer draws on average instead of a
scan of length $x$. Coupling two counts estimates $\ln(a/b)$ without
$\gamma$, and a binary range reduction then bounds the standard error
by $\sqrt{\ln 2/n}$ for every $x$ at fewer than $1.7$ draws per
repetition on average. A classical deterministic series on the same reduction reaches the
rounding floor with sixteen terms in the reported experiments,
quantifying the gap between the two approaches. The purpose is
pedagogical: the analysis of an algorithm becomes an estimator of the
quantity that analysis produces, with exact variances, finite-sample
error bounds, reproducible experiments, and machine-checked algorithm
contracts and moment identities. The ingredients are classical; the
contribution is their assembly.
\end{abstract}

\section{Motivation}\label{sec:intro}

A standard exercise in the analysis of algorithms asks how many times
the running maximum changes when a list of $x$ distinct numbers in
random order is scanned once (Algorithm~\ref{alg:max}). The answer is
the harmonic number $H_x = \sum_{i=1}^{x} 1/i$ on average, which is
$\ln x + \gamma + O(1/x)$ with $\gamma \approx 0.5772$ the
Euler--Mascheroni constant. The exercise appears in Knuth's analysis
of his Algorithm~M for finding the maximum~\cite{Knuth97} and, in the
guise of the hiring problem, in Cormen, Leiserson, Rivest, and
Stein~\cite{CLRS09}. The replacement events are the \emph{records} of
the sequence, whose theory goes back to R\'enyi~\cite{Renyi62}; see
Arnold, Balakrishnan, and Nagaraja~\cite{ABN98} for a monograph
treatment.

\begin{algorithm}[ht]
\caption{\textsc{Compute-Max}$(L[1..x])$}\label{alg:max}
\begin{algorithmic}[1]
\State $\mathit{max} \gets -\infty$
\For{$i \gets 1$ \textbf{to} $x$} \Comment{$x$ comparisons}
  \If{$L[i] > \mathit{max}$}
    \State $\mathit{max} \gets L[i]$ \Comment{executed $R_x$ times}
  \EndIf
\EndFor
\State \Return $\mathit{max}$
\end{algorithmic}
\end{algorithm}

This note runs the exercise backwards. If we generate the list
ourselves from a random number generator, we can count the
replacements, average the count over $n$ independent lists, and read
off an estimate of $H_x$ and then of $\ln x$, with a one-line change
to a loop every programmer has written. The inner loop does comparison
and counting only, and the algorithm computes, by sampling, the very
function that appears in its own running-time analysis. The
construction is a teaching example, and we draw three lessons from it: how the analysis
of an algorithm becomes an estimator with an exact variance, honest
error bars, and a separation of sampling error from deterministic
bias (Sections~\ref{sec:scan} and~\ref{sec:accuracy}); how the same
analysis yields cheaper samplers and lower variance, from a scan of
length $x$ to $H_x$ integer draws on average and then to a uniformly
bounded expected number of draws for every $x$
(Sections~\ref{sec:sampler} and~\ref{sec:reduction}); and how Monte Carlo convergence at rate
$n^{-1/2}$ compares with the geometric convergence of a deterministic
series (Section~\ref{sec:series}).

The classical ingredients are cited where they appear: R\'enyi's
independence of the record indicators, the position-of-the-maximum
recursion behind discrete stick breaking, Bernstein's inequality, and
the binary reduction and transformed series of library logarithms.
The contribution is their assembly into one worked example with exact
variances and finite-sample error bounds, a reproducible experimental
record, corrections to the 2017 version, and machine-checked proofs
of the algorithms' contracts and of the finite-sum identities behind
the error bounds. Our experiments (Section~\ref{sec:exp-series})
compare the randomized estimators with deterministic logarithm
computation: one randomized estimate of $\ln 384$ with a thousand
repetitions takes about a millisecond and has root mean squared error
about $0.02$, while sixteen terms of a classical series after the same
range reduction take about $2.5$ microseconds and reach the rounding
floor of double precision. The 2017 version of this note~\cite{Dasdan17}
contained two incorrect statements and one that invited misreading;
Appendix~\ref{app:revision} records the corrections.

\section{Records and the Scan Estimator}\label{sec:scan}

Let $U_1, \dots, U_x$ be independent and identically distributed
random variables with a continuous distribution, so that ties have
probability zero. (Uniform on $[0,1)$ is convenient but not
necessary; a uniformly random permutation of $x$ distinct values gives
the same results.) Define the record indicators
\begin{equation}\label{eq:indicator}
  I_1 = 1, \qquad
  I_i = \mathbf{1}\{U_i > \max(U_1, \dots, U_{i-1})\}, \quad 2 \leq i \leq x,
\end{equation}
where $\mathbf{1}\{\cdot\}$ is the indicator function, equal to $1$
when the event in braces occurs and to $0$ otherwise, and the record
count $R_x = \sum_{i=1}^{x} I_i$. Then $R_x$ is exactly
the number of times line~4 of Algorithm~\ref{alg:max} executes on the
input $L = (U_1, \dots, U_x)$.

\begin{proposition}[R\'enyi~{\cite[Lemma~1, p.~8]{Renyi62}}]\label{prop:records}
The indicators $I_1, \dots, I_x$ are independent with
$\Pr(I_i = 1) = 1/i$. Consequently
\begin{equation}\label{eq:meanvar}
  \E R_x = H_x,
  \qquad
  \Var R_x = \sum_{i=1}^{x} \frac{1}{i}\Bigl(1 - \frac{1}{i}\Bigr)
           = H_x - H^{(2)}_x,
\end{equation}
where $H^{(2)}_x = \sum_{i=1}^{x} 1/i^2$. (The literature writes
$H_n$; here $n$ is reserved for the number of repetitions.)
\end{proposition}

\begin{proof}
By exchangeability of $U_1, \dots, U_i$, each of the $i!$ relative
orders of the first $i$ values is equally likely, so $U_i$ is the
largest of the first $i$ with probability $1/i$; this gives
$\Pr(I_i = 1) = 1/i$. Let $\rho_i \in \{1, \dots, i\}$ be the rank of
$U_i$ among $U_1, \dots, U_i$. The map from relative orders of
$(U_1, \dots, U_x)$ to $(\rho_1, \dots, \rho_x)$ is a bijection
between the $x!$ permutations and the $x!$ tuples with $\rho_i \leq i$,
so the $\rho_i$ are independent and uniform. Since $I_i =
\mathbf{1}\{\rho_i = i\}$, the $I_i$ are independent. The mean and
variance follow by summing.
\end{proof}

The argument needs the prefix statement ``$U_i$ is the largest of the
\emph{first $i$} values with probability $1/i$'' for every $i$; the
statement ``each value is the overall maximum with probability $1/x$''
accounts only for the last replacement, and independence, which the
rank argument supplies, is what makes the variance a sum.

\begin{lemma}[Elementary variance bounds]\label{lem:variance-bounds}
For every integer $x \geq 1$,
\begin{equation}\label{eq:variance-bounds}
  \max\!\left\{0,\,\ln x + \gamma - \frac{\pi^2}{6}\right\}
  \leq \Var R_x \leq \ln x.
\end{equation}
\end{lemma}

\begin{proof}
$H_x > \ln x + \gamma$, since $H_x - \ln x$ decreases strictly to
$\gamma$, and $H^{(2)}_x < \sum_{i=1}^{\infty} i^{-2} = \pi^2/6$; these
give the lower bound, together with the nonnegativity of variance. For
the upper bound,
\[
  \Var R_x = H_x - H^{(2)}_x
  \leq H_x - 1
  = \sum_{i=2}^{x} \frac{1}{i}
  \leq \int_1^x \frac{dt}{t} = \ln x.
\qedhere
\]
\end{proof}

Thus the variance differs from $\ln x$ by at most $\pi^2/6 - \gamma
\approx 1.0677$, and the standard error of the average of $n$
independent record counts is at most $\sqrt{\ln x/n}$, without
evaluating either harmonic sum.

Algorithm~\ref{alg:scan} repeats the scan $n$ times on fresh random
lists and averages the counts. It is the algorithm of the 2017
version with three changes: the sample variance uses $n-1$, the
correction term is improved (Section~\ref{sec:accuracy}), and
$\ln 1 = 0$ is returned exactly.

\begin{algorithm}[ht]
\caption{\textsc{Scan-Estimate}$(x, n)$: estimate $H_x$ and $\ln x$ by
  $n \geq 2$ scans of length $x \geq 1$}\label{alg:scan}
\begin{algorithmic}[1]
\For{$r \gets 1$ \textbf{to} $n$}
  \State $c_r \gets 0$; $\mathit{max} \gets -\infty$
  \For{$i \gets 1$ \textbf{to} $x$}
    \State $u \gets \textsc{Uniform}[0,1)$
    \If{$u > \mathit{max}$}
      \State $\mathit{max} \gets u$; $c_r \gets c_r + 1$
    \EndIf
  \EndFor
\EndFor
\State $\Hbar \gets \frac{1}{n}\sum_{r=1}^{n} c_r$
  \Comment{estimate of $H_x$}
\State $s^2 \gets \frac{1}{n-1}\sum_{r=1}^{n} (c_r - \Hbar)^2$
  \Comment{sample variance; requires $n \geq 2$}
\State $\mathit{se} \gets \sqrt{s^2/n}$
  \Comment{estimated standard error of $\Hbar$}
\If{$x = 1$} \Return $(\Hbar, 0, \mathit{se})$ \EndIf
\State $c(x) \gets \dfrac{1}{2x} - \dfrac{1}{12x^2}$
\State \Return $(\Hbar,\; \Hbar - \gamma - c(x),\; \mathit{se})$
  \Comment{$(\hat H_x,\; \Lhat_x,\; \mathit{se})$}
\end{algorithmic}
\end{algorithm}

Each $c_r$ is an independent copy of $R_x$, so $\Hbar$ is unbiased for
$H_x$ with $\Var \Hbar = (H_x - H^{(2)}_x)/n \leq (\ln x)/n$ by
Lemma~\ref{lem:variance-bounds}; that $R_x$ is an integer while $H_x$
is not for $x > 1$~\cite{SoWe07} is no obstacle, since integer-valued
random variables routinely have non-integer means. The analysis
assumes independent, exactly uniform draws from the indicated ranges.
A floating-point implementation of the scan additionally permits ties:
for independent draws uniform on a grid of $2^{53}$ values, each pair
ties with probability $2^{-53}$, negligible for the scan sizes
considered here. The integer sampler of Section~\ref{sec:sampler}
avoids the tie issue, provided each bounded-integer draw is
conditionally uniform given all previous draws. Pseudorandom
implementations approximate the ideal randomness assumed by the
analysis.

\section{Exact Sampling with \texorpdfstring{$H_x$}{H\_x} Draws on Average}\label{sec:sampler}

Algorithm~\ref{alg:scan} spends $x$ draws and comparisons per
repetition to produce one integer whose mean is about $\ln x$. The
analysis that gives the mean also gives a shortcut: the distribution of
$R_x$ can be sampled exactly with about $\ln x$ draws, without
generating the list and without computing a maximum.

\begin{algorithm}[ht]
\caption{\textsc{Sample-Records}$(x)$: exact sample of $R_x$}\label{alg:sampler}
\begin{algorithmic}[1]
\State $c \gets 0$; $m \gets x$
\While{$m > 0$}
  \State $c \gets c + 1$
  \State $m \gets \textsc{Uniform}\{0, 1, \dots, m-1\}$
    \Comment{one integer draw}
\EndWhile
\State \Return $c$
\end{algorithmic}
\end{algorithm}

\begin{proposition}\label{prop:sampler}
The output of Algorithm~\ref{alg:sampler} has the distribution of
$R_x$ in~\eqref{eq:indicator}. The number of integer draws equals the
output, so its expectation is $H_x$ and its variance is $H_x -
H^{(2)}_x$.
\end{proposition}

\begin{proof}
Let $J$ be the position of the maximum of $U_1, \dots, U_x$. By
exchangeability, $J$ is uniform on $\{1, \dots, x\}$. Position $J$ is
a record, no position after $J$ is a record, and the records before
$J$ are exactly the records of the prefix $U_1, \dots, U_{J-1}$.
Conditioned on $J$, the relative order of that prefix is a uniformly
random permutation of $J-1$ elements. Hence, with $R_0 = 0$,
\begin{equation}\label{eq:recurrence}
  R_x \overset{d}{=} 1 + R'_{J-1}, \qquad J \sim \textsc{Uniform}\{1, \dots, x\},
\end{equation}
where $\overset{d}{=}$ denotes equality in distribution and $R'$ is a
copy of the record-count process independent of $J$. Algorithm~\ref{alg:sampler}
implements~\eqref{eq:recurrence} iteratively: each pass through the
loop adds one record and replaces $m$ by $J - 1$, which is uniform on
$\{0, \dots, m-1\}$. The loop runs once per record, so the number of
draws is the output. The moments are those of
Proposition~\ref{prop:records}.
\end{proof}

Algorithm~\ref{alg:sampler} is the counting version of classical
\emph{discrete uniform stick breaking}: when $m$ units remain, choose
$K$ uniformly from $\{1, \dots, m\}$ and replace $m$ by $m - K$ until
zero. The construction appears in the permutation-cycle problem solved
by Lossers~\cite{Lossers10} and in Abramson et
al.~\cite[Definition~8 and Proposition~9]{APRU11}; since $m - K$ is
uniform on $\{0, \dots, m-1\}$, counting its steps gives exactly the
transition rule of Algorithm~\ref{alg:sampler}, and the selected sizes
are the cycle lengths of a uniform random permutation, whose count is
equal in distribution to the record count. We use its record
interpretation.
Devroye~\cite[Chapter~XIV, Section~1]{Devroye86} gives a different,
forward method that generates the record \emph{times} of an i.i.d.\
sequence with logarithmic expected work by jumping over non-records
with geometric steps; Algorithm~\ref{alg:sampler} works backwards from
the maximum and produces only the count. The cost model counts one uniform integer
draw as unit cost; the bit complexity of a draw from $\{0, \dots,
m-1\}$ grows with $\ln m$, so the expected bit cost is $O(\ln^2 x)$
rather than $O(\ln x)$. Replacing the inner loop of
Algorithm~\ref{alg:scan} by a call to Algorithm~\ref{alg:sampler}
leaves every output distribution unchanged and reduces the expected
cost of a repetition from $x$ to $H_x$. It also makes the 2017
version's description ``integer arithmetic only'' literal: the scan's
arithmetic is integer counting but its comparisons are between real
numbers, whereas the sampler draws and compares integers only
(Appendix~\ref{app:revision}).

\section{Bias, Variance, and Error Guarantees}\label{sec:accuracy}

\begin{lemma}\label{lem:bias}
For every integer $x \geq 1$,
\begin{equation}\label{eq:expansion}
  H_x = \ln x + \gamma + c(x) + r_x,
  \qquad
  c(x) = \frac{1}{2x} - \frac{1}{12x^2},
  \qquad
  0 < r_x < \frac{1}{120\,x^4},
\end{equation}
and $r_{x+1} < r_x$.
\end{lemma}

The proof, from the asymptotic expansion of the digamma function and
an elementary identity for $r_x - r_{x+1}$, is in
Appendix~\ref{app:proofs}. The 2017 version used the correction
$1/(2x)$ alone, whose error is of order $1/(12x^2)$, about $0.005$ at
$x = 4$ and not negligible against the sampling error there; with
$c(x)$ the deterministic error is below $3.3 \times 10^{-5}$ already
at $x = 4$ and below $5 \times 10^{-10}$ for $x \geq 64$.

\subsection{Sampling error and bias}

Let $\Hbar$ be the average of $n$ independent copies of $R_x$ and, for
$x \geq 2$, $\Lhat_x = \Hbar - \gamma - c(x)$. By
Lemma~\ref{lem:bias},
\begin{equation}\label{eq:decomp}
  \Lhat_x - \ln x = (\Hbar - H_x) + r_x,
\end{equation}
so the estimator has bias $r_x \in (0, 1/(120x^4))$, which no number
of repetitions removes, and Monte Carlo error with
\begin{equation}\label{eq:mse}
  \E\bigl(\Lhat_x - \ln x\bigr)^2
  = \frac{H_x - H^{(2)}_x}{n} + r_x^2
  \leq \frac{\ln x}{n} + \frac{1}{14400\,x^8},
\end{equation}
by Lemma~\ref{lem:variance-bounds} and Lemma~\ref{lem:bias}. In
floating point the stored $\gamma$ adds a further deterministic error
of $-4.9 \times 10^{-18}$, which exceeds $r_x$ in magnitude beyond
$x \approx 6400$, and the final arithmetic adds rounding on the scale
of $2^{-53} \ln x$; both are far below the sampling error at any
feasible $n$ (Appendix~\ref{app:proofs}).

\subsection{How the error scales}

By Lemma~\ref{lem:variance-bounds}, the standard error
$\operatorname{SE}$ of $\Lhat_x$ satisfies, for every $x > 1$,
\begin{equation}\label{eq:se}
  \operatorname{SE}(\Lhat_x) = \sqrt{\frac{H_x - H^{(2)}_x}{n}}
  \leq \sqrt{\frac{\ln x}{n}},
\end{equation}
and since $H_x - H^{(2)}_x = \ln x + \gamma - \pi^2/6 + O(1/x)$, the
ratio of this bound to the exact standard error tends to one as
$x \to \infty$. The
absolute standard error thus \emph{grows} slowly with $x$, while the
relative standard error $\operatorname{SE}(\Lhat_x)/\ln x \leq
1/\sqrt{n \ln x}$ ($x \geq 2$) decreases: larger arguments are easier
only relatively, which is the pattern seen in the 2017 version's
figure, where the relative error fell with $x$.
Section~\ref{sec:reduction} removes the growth of the bound.

For a target absolute error $\varepsilon$ at confidence $1 - \delta$,
the central limit theorem suggests the approximate planning rule
$n \approx z_{1-\delta/2}^2 (H_x - H^{(2)}_x)/\varepsilon^2$, or
$n \geq z_{1-\delta/2}^2 \ln x/\varepsilon^2$ with the bound in place
of the variance, both assuming a suitable normal approximation and
negligible bias; for $\varepsilon = 0.01$ and $\delta = 0.05$ at
$x = 256$ these give $n \approx 1.7 \times 10^5$ and $2.1 \times
10^5$.

\subsection{A finite-sample rule}\label{sec:bernstein}

A guarantee valid at every $n$ follows from Bernstein's
inequality~\cite[Theorem~2.10 and Corollary~2.11]{BLM13}.
Let an estimate $L$ of a target $\theta$ satisfy
\begin{equation}\label{eq:error-split}
  L - \theta = S + \beta,
\end{equation}
where $S$ is the centered sampling term, the average over $n$
independent repetitions of a sum of independent indicators minus its
mean, with per-repetition variance at most $V$, and $\beta$ is a
deterministic bias with $|\beta| \leq B$. For $\Lhat_x$,
by~\eqref{eq:decomp}, $S = \Hbar - H_x$ and $\beta = r_x$. Let
$\varepsilon > B$ and $\delta \in (0, 1)$. The error reaches
$\varepsilon$ only if the sampling term reaches $t = \varepsilon - B$,
and $S$ is $1/n$ times the sum, over all $n$ repetitions and all their
indicators, of independent centered Bernoulli variables bounded by $1$
with total variance at most $nV$, so
\begin{equation}\label{eq:bernstein}
  \Pr\bigl(|L - \theta| \geq \varepsilon\bigr)
  \leq \Pr\bigl(|S| \geq t\bigr)
  \leq 2 \exp\!\left(-\frac{n t^2}{2\,(V + t/3)}\right),
\end{equation}
so that
\begin{equation}\label{eq:sample-size}
  n \geq \left\lceil
    \frac{2\,(V + t/3)}{t^2}\,\ln\frac{2}{\delta}
  \right\rceil
\end{equation}
repetitions suffice for absolute error below $\varepsilon$ with
probability at least $1 - \delta$.
For $\Lhat_x$, with $S = \Hbar - H_x$, take $V = H_x - H^{(2)}_x$ or
$V = \ln x$ and $B = 1/(120 x^4)$: for $\varepsilon = 0.01$, $\delta = 0.05$, $x = 256$
this gives $n = 331\,014$ with the exact variance and $409\,356$ with
$\ln x$, about $1.9$ times the normal-approximation values, the price
of a guarantee rather than an approximation. Section~\ref{sec:reduction}
instantiates the same rule with $V = \ln 2$. We use $n$ fixed in
advance; stopping when a running confidence interval first looks
narrow enough does not preserve its coverage.

\section{Coupling and Range Reduction}\label{sec:reduction}

The bound in~\eqref{eq:se} grows with $x$ because the estimator
samples all of $\ln x$. Two steps remove the growth: estimating a
difference of logarithms from one coupled count, and reducing the
argument to the interval $[1, 2)$ by its binary representation. The
bounds of this section concern the ideal sampling error; the
implemented error has further parts (Remark~\ref{rem:ln2}).

\subsection{Coupled tail counts}\label{sec:rational}

For integers $a > b \geq 1$ let $T_{a,b} = \sum_{i=b+1}^{a} I_i$ be
the number of records at positions $b+1, \dots, a$ of a single
sequence. By Proposition~\ref{prop:records},
\begin{equation}\label{eq:tail}
  \E T_{a,b} = H_a - H_b,
  \qquad
  \Var T_{a,b} = \sum_{i=b+1}^{a} \Bigl(\frac{1}{i} - \frac{1}{i^2}\Bigr),
\end{equation}
and by Lemma~\ref{lem:bias}
\begin{equation}\label{eq:lnab}
  \ln\frac{a}{b} = (H_a - H_b) - c(a) + c(b) - r_a + r_b .
\end{equation}
So if $\bar T$ is the average of $n$ copies of $T_{a,b}$, the
estimator $\bar T - c(a) + c(b)$ of $\ln(a/b)$ has bias $r_a - r_b$,
which lies in $(-1/(120b^4), 0)$ for $a > b$ by the positivity and
strict decrease of the remainders in Lemma~\ref{lem:bias}, and is zero
when $a = b$; the constant $\gamma$ does not appear. The variance is
$\Var T_{a,b}/n$, and as in Lemma~\ref{lem:variance-bounds},
$\Var T_{a,b} \leq \sum_{i=b+1}^{a} 1/i \leq \ln(a/b)$, so the
standard error is at most $\sqrt{\ln(a/b)/n}$. Subtracting two
independent estimates of $H_a$ and $H_b$ instead would have variance
$(\Var R_a + \Var R_b)/n$, exceeding the coupled variance $(\Var R_a -
\Var R_b)/n$ by $2\Var R_b/n$, which is zero for $b = 1$ and positive
for $b \geq 2$; coupling is most valuable when $a$ and $b$ are close.
For $b = 1$ and large $a$ the bias is $r_a - r_1 \approx -0.0061$:
$\gamma$ cancels, but the correction at $b = 1$ is crude. Scaling
$(a, b)$ to $(ka, kb)$ for an integer $k \geq 1$ leaves $\ln(a/b)$
unchanged, shrinks the bias bound by $k^{-4}$, and keeps the expected
number of draws below $1 + \ln(a/b)$, at the price of drawing from
larger ranges.

\begin{algorithm}[ht]
\caption{\textsc{Sample-Tail-Records}$(a, b)$: exact sample of $T_{a,b}$,
  $a \geq b \geq 1$}\label{alg:tail}
\begin{algorithmic}[1]
\If{$a = b$} \Return $0$ \EndIf
\State $c \gets 0$; $m \gets a$
\Loop
  \State $J \gets \textsc{Uniform}\{1, \dots, m\}$
  \If{$J \leq b$} \Return $c$ \EndIf
  \State $c \gets c + 1$; $m \gets J - 1$
\EndLoop
\end{algorithmic}
\end{algorithm}

\begin{proposition}\label{prop:tail}
Algorithm~\ref{alg:tail} returns a variable distributed as $T_{a,b}$.
For $a > b$ it uses exactly $c + 1$ draws, so $1 + H_a - H_b$ draws on
average; for $a = b$ it uses none.
\end{proposition}

\begin{proof}
Let $J$ be the position of the maximum of $U_1, \dots, U_m$, uniform
on $\{1, \dots, m\}$. If $J \leq b$, no position in $b+1, \dots, m$
can be a record, since a record there would exceed $U_J$. If $J > b$,
position $J$ is a record in the tail, positions $J+1, \dots, m$ are
not, and the records in $b+1, \dots, J-1$ are the tail records of the
prefix of length $J - 1$, whose relative order is again uniform. The
loop implements this recursion and stops at the first draw with
$J \leq b$, which happens exactly once. For $a = b$, $T_{a,b} = 0$
and the algorithm returns $0$ directly.
\end{proof}

The normal approximation behind an interval $\bar T \pm 1.96
\sqrt{s^2/n}$ fails when records in the tail are rare. For $(a, b) =
(1001, 1000)$ the count is Bernoulli$(1/1001)$, so with $n = 1000$ all
counts are zero with probability $(1000/1001)^{1000} \approx 0.368$;
the sample variance is then zero and the interval is a point that
misses the target. Exact binomial calculation gives coverage $0.63$
for that interval rather than $0.95$ (Section~\ref{sec:exp-theory}).
The Bernstein rule~\eqref{eq:bernstein}, with $S = \bar T - \E \bar T$,
$\beta = r_a - r_b$, and $V = \Var T_{a,b}$, remains valid in this
regime.

\subsection{Binary range reduction}

For an integer $x \geq 1$ let $k$ be the number of bits of $x$ minus
one and $b = 2^k$, so that $b \leq x < 2b$; $k$ is the integer part of
$\log_2 x$, read off the binary representation by a bit-length
operation or a shift. Then
\begin{equation}\label{eq:reduce}
  \ln x = k \ln 2 + \ln\frac{x}{b},
  \qquad 0 \leq \ln\frac{x}{b} < \ln 2 .
\end{equation}
The first term is exact given the constant $\ln 2$; only the second,
which lies in $[0, \ln 2)$, is estimated, by the coupled estimator
with $(a, b)$ replaced by $(qx, qb)$ for an integer $q \geq 1$, which
is allowed because $\ln(qx/qb) = \ln(x/b)$. This is the standard
argument reduction of deterministic logarithm routines, which write
$x = 2^k(1 + f)$ and reconstruct $k \ln 2 + \ln(1 + f)$~\cite{fdlibm}.
Algorithm~\ref{alg:reduced} is the result; read $q = 1$ on a first
pass, since $q$ is a bias control explained below. Any positive
integer $d \leq x$ with known $\ln d$ would serve as the reference
point; powers of
two are the natural choice because $k$ and $b$ come from the bits of
$x$. The constant $\gamma$ does not appear; the constant $\ln 2$ does
(Remark~\ref{rem:ln2}).

\begin{algorithm}[ht]
\caption{\textsc{Reduced-Estimate}$(x, n, q)$: estimate $\ln x$ with
  standard error at most $\sqrt{\ln 2/n}$; $x \geq 1$, $n \geq 2$,
  $q \geq 1$}\label{alg:reduced}
\begin{algorithmic}[1]
\State $k \gets (\text{number of bits of } x) - 1$; $b \gets 2^k$
  \Comment{$b \leq x < 2b$; a shift, not a logarithm}
\If{$x = b$} \Return $(k \ln 2,\; 0)$ \EndIf
  \Comment{powers of two, including $x = 1$: exact, no draws}
\For{$r \gets 1$ \textbf{to} $n$}
  \State $t_r \gets \textsc{Sample-Tail-Records}(qx, qb)$
    \Comment{Algorithm~\ref{alg:tail}; $t_r + 1$ integer draws}
\EndFor
\State $\bar T \gets \frac{1}{n}\sum_{r=1}^{n} t_r$
  \Comment{estimate of $H_{qx} - H_{qb}$}
\State $s^2 \gets \frac{1}{n-1}\sum_{r=1}^{n} (t_r - \bar T)^2$
  \Comment{sample variance}
\State \Return $\bigl(k \ln 2 + \bar T - c(qx) + c(qb),\; \sqrt{s^2/n}\bigr)$
  \Comment{$(\Lred_x,\; \mathit{se})$}
\end{algorithmic}
\end{algorithm}

\begin{proposition}\label{prop:reduced}
Let $\Lred_x$ be the first output of Algorithm~\ref{alg:reduced}, with
exact arithmetic and exact $\ln 2$. If $x$ is a power of two,
$\Lred_x = \ln x$. Otherwise its bias is strictly negative, and
\begin{equation}\label{eq:reduced-bias}
  \E \Lred_x - \ln x = r_{qx} - r_{qb},
  \qquad
  \bigl|\E \Lred_x - \ln x\bigr| < \frac{1}{120\,(qb)^4}
  \leq \frac{1}{120\,q^4},
\end{equation}
\begin{equation}\label{eq:reduced-var}
  \Var \Lred_x
  = \frac{1}{n} \sum_{i=qb+1}^{qx} \Bigl(\frac{1}{i} - \frac{1}{i^2}\Bigr)
  \leq \frac{\ln(x/b)}{n} < \frac{\ln 2}{n},
\end{equation}
and each repetition uses $t_r + 1$ integer draws, so
$1 + H_{qx} - H_{qb} \leq 1 + \ln(x/b) < 1 + \ln 2$ on average.
\end{proposition}

\begin{proof}
By~\eqref{eq:tail} and~\eqref{eq:lnab} with $(a, b)$ replaced by
$(qx, qb)$, $\E \bar T - c(qx) + c(qb) = \ln(x/b) + r_{qx} - r_{qb}$;
adding $k \ln 2$ and using~\eqref{eq:reduce} gives the expectation.
By Lemma~\ref{lem:bias}, $0 < r_{qx} < r_{qb} < 1/(120(qb)^4)$ because
$qb < qx$, which gives the sign and the bias bound, and $b \geq 1$
gives the second inequality. The variance is $\Var T_{qx,qb}/n$
by~\eqref{eq:tail}, with $\sum_{i=qb+1}^{qx} 1/i \leq \int_{qb}^{qx}
dt/t = \ln(x/b) < \ln 2$ as in Section~\ref{sec:rational}. The draw
count is Proposition~\ref{prop:tail} with $H_{qx} - H_{qb} \leq
\ln(x/b)$ by the same integral bound.
\end{proof}

Consequently
\begin{equation}\label{eq:reduced-mse}
  \E\bigl(\Lred_x - \ln x\bigr)^2
  = \Var \Lred_x + (r_{qx} - r_{qb})^2
  \leq \frac{\ln(x/b)}{n} + \frac{1}{14400\,(qb)^8}
  \leq \frac{\ln 2}{n} + \frac{1}{14400\,q^8},
\end{equation}
a bound that does not depend on $x$, whereas~\eqref{eq:mse} grows
without limit. The magnitude of the bias is bounded in terms of
$b \geq x/2$ alone; increasing $q$ reduces the bias bound while the
variance per repetition and the expected draws increase towards their
limits $\ln(x/b)$ and $1 + \ln(x/b)$, within the same uniform bounds,
and the integers handled have about $\log_2(qx)$ bits
(Table~\ref{tab:e6d} in Appendix~\ref{app:tables}).

\subsection{Uniform bounds, sample size, and cost}

By~\eqref{eq:reduced-var}, the standard error $\operatorname{SE}$ of
$\Lred_x$ satisfies, for every $x \geq 1$,
\begin{equation}\label{eq:reduced-se}
  \operatorname{SE}(\Lred_x) \leq \sqrt{\frac{\ln(x/b)}{n}}
  < \sqrt{\frac{\ln 2}{n}},
\end{equation}
so the bound on the absolute standard error is uniform in $x$, and for
$x \geq 2$ the relative standard error is below $\sqrt{\ln 2/n}/\ln x$,
which falls like $1/\ln x$ rather than the $1/\sqrt{\ln x}$
of~\eqref{eq:se}. Within the bound, the exact standard error depends
on the position of $x$ in $[b, 2b)$: zero at powers of two, small just
above them, largest just below the next power of two.

The planning rules of Section~\ref{sec:accuracy} become independent
of $x$. The normal approximation with $V = \ln 2$ gives $n \geq
z_{1-\delta/2}^2 \ln 2/\varepsilon^2$, about $2.7 \times 10^4$ for
$\varepsilon = 0.01$ and $\delta = 0.05$, against $1.7 \times 10^5$
for $\Lhat_x$ at $x = 256$. The Bernstein rule~\eqref{eq:bernstein}
applies with $S = \bar T - \E \bar T$ and $\beta = r_{qx} - r_{qb}$
in~\eqref{eq:error-split}, $V = \ln 2$ (the count $T_{qx,qb}$ is a sum
of independent indicators), and the uniform bias allowance $B = 1/(120
q^4)$, or the sharper $1/(120 (qb)^4)$ for a given $x$; the rule
also tells how to choose $q$: $B$ must lie below $\varepsilon$, and
well below it to avoid an excessive $n$. With $q = 1$ the uniform
allowance $1/120 \approx 0.0083$ consumes most of a budget
$\varepsilon = 0.01$, while $q = 4$ gives $B \approx 3.3 \times
10^{-5}$ and $n = 51\,721$ for every $x$ at $\delta = 0.05$ (with the
requirement $n \geq 2$ of Algorithm~\ref{alg:reduced}). A coarse
allowance that fails does not show that $q = 1$ cannot meet a
tolerance, since the sharper argument-specific $B$ may suffice. These
are absolute tolerances, and the exact-arithmetic, exact-$\ln 2$
qualification of Remark~\ref{rem:ln2} remains in force.

The cost falls as well: for $x > b$ a repetition uses $1 + H_{qx} -
H_{qb} < 1 + \ln 2 \approx 1.69$ integer draws on average, and none
for $x = b$, against $H_x \approx \ln x$ for Algorithm~\ref{alg:sampler}
and $x$ real draws for the scan, plus constant deterministic work.
Judged by variance times expected draws per repetition, the reduced
estimator costs at most $\ln 2\,(1 + \ln 2) < 1.18$ for every $x$ and
$q$, while the original costs $(H_x - H^{(2)}_x) H_x$, which grows
with $x$; the reduced estimator is the more efficient for every
$x \geq 2$ in this unit-draw model (Appendix~\ref{app:proofs}).

\begin{remark}[The constant $\ln 2$ and floating point]\label{rem:ln2}
Algorithm~\ref{alg:reduced} needs $\ln 2$ as Algorithm~\ref{alg:scan}
needs $\gamma$, and multiplies it by $k$. With a stored approximation
$\ell_2$ every estimate acquires the deterministic error $k(\ell_2 -
\ln 2)$, about $-2.3 \times 10^{-17}\,k$ for the double-precision
value used in our experiments, and the final arithmetic adds rounding
on the scale of $2^{-53} \ln x$. These are negligible against the
sampling error in the configurations studied, but not in general: the
sampling error is zero at powers of two and arbitrarily small just
above them, and the constant's term grows with $k$. The uniform
bounds~\eqref{eq:reduced-var}--\eqref{eq:reduced-se} are statements
about the ideal model with exact $\ln 2$, not about floating-point
arithmetic at a fixed precision for arbitrarily large $x$; computing
$\ln 2$ to the precision required is separate deterministic work,
exactly as for $\gamma$ (Appendix~\ref{app:proofs}).
\end{remark}

\section{A Deterministic Comparison}\label{sec:series}

The reduction of Section~\ref{sec:reduction} is also the first step of
the classical deterministic algorithms for the logarithm, and
completing it deterministically makes the comparison with the
randomized estimators concrete. With $b \leq x < 2b$ as before, put
$m = x/b \in [1, 2)$ and
\begin{equation}\label{eq:zdef}
  z = \frac{m - 1}{m + 1} = \frac{x - b}{x + b},
  \qquad 0 \leq z < \frac{1}{3}.
\end{equation}
Then $m = (1 + z)/(1 - z)$ and
\begin{equation}\label{eq:series}
  \ln x = k \ln 2 + \ln\frac{1 + z}{1 - z}
        = k \ln 2 + 2 \sum_{j \geq 0} \frac{z^{2j+1}}{2j + 1},
\end{equation}
the series of the inverse hyperbolic tangent.
Algorithm~\ref{alg:series} truncates the sum after $N$ terms and
evaluates it by Horner's rule in the variable $z^2$.

\begin{algorithm}[ht]
\caption{\textsc{Series-Log}$(x, N)$: deterministic $\ln x$ after binary
  reduction; integers $x \geq 1$ and $N \geq 1$}\label{alg:series}
\begin{algorithmic}[1]
\State $k \gets (\text{number of bits of } x) - 1$; $b \gets 2^k$
  \Comment{as in Algorithm~\ref{alg:reduced}}
\If{$x = b$} \Return $k \ln 2$ \EndIf
\State $z \gets (x - b)/(x + b)$; \quad $w \gets z^2$; \quad $s \gets 0$
\For{$j \gets N - 1$ \textbf{down to} $0$}
  \State $s \gets s \cdot w + 1/(2j + 1)$
\EndFor
\State \Return $k \ln 2 + 2 z s$
  \Comment{at loop exit $s = \sum_{j < N} w^{j}/(2j+1)$ (Horner)}
\end{algorithmic}
\end{algorithm}

\begin{proposition}\label{prop:series}
Let $L_N$ be the output of Algorithm~\ref{alg:series} in exact
arithmetic with exact $\ln 2$. If $x$ is a power of two, $L_N = \ln x$.
Otherwise
\begin{equation}\label{eq:series-bound}
  0 < \ln x - L_N
  \leq \frac{2 z^{2N+1}}{(2N+1)(1 - z^2)}
  \leq \frac{9}{4\,(2N+1)\,3^{2N+1}} .
\end{equation}
\end{proposition}

\begin{proof}
For $|z| < 1$, $\ln\frac{1+z}{1-z} = \int_0^z \frac{2\,dt}{1 - t^2} =
2 \sum_{j \geq 0} \frac{z^{2j+1}}{2j+1}$, by expanding the integrand in
a geometric series and integrating term by term;
with~\eqref{eq:zdef} this is~\eqref{eq:series}. The tail after $N$
terms, $2 \sum_{j \geq N} z^{2j+1}/(2j+1)$, is positive for $z > 0$
and at most $\frac{2}{2N+1} \sum_{j \geq N} z^{2j+1} = \frac{2
z^{2N+1}}{(2N+1)(1 - z^2)}$. Since $z < 1/3$, $z^{2N+1} <
3^{-(2N+1)}$ and $1 - z^2 > 8/9$, which gives the last bound. For $x =
b$, $z = 0$ and $L_N = k \ln 2 = \ln x$.
\end{proof}

The uniform bound is approximately $1.9 \times 10^{-3}$ for $N = 2$,
$1.3 \times 10^{-5}$ for $N = 4$, $1.03 \times 10^{-9}$ for $N = 8$,
and $1.23 \times 10^{-17}$ for $N = 16$. It decreases geometrically,
by a factor of about nine per term, so $N = O(\log(1/\varepsilon))$
terms give truncation error $\varepsilon$, independent of $x$ after
the reduction, at $O(N)$ arithmetic operations (an operation count,
not a bit complexity). The randomized estimators need
$O(\varepsilon^{-2})$ repetitions for the same absolute tolerance at a
fixed confidence once the bias is controlled
(Sections~\ref{sec:accuracy} and~\ref{sec:reduction}), and at fixed
$x$ and $q$ otherwise have a bias floor.

The implemented error again has three parts: the truncation
error~\eqref{eq:series-bound}, the stored-constant term $k(\ell_2 -
\ln 2)$, and the rounding of the arithmetic, on the scale of $2^{-53}
\ln x$. With $16$ terms the truncation bound is about $1.3 \times
10^{-17}$ and the other two parts dominate
(Section~\ref{sec:exp-series}); a truncation bound of that size is
therefore not a total binary64 error of that size, and a tolerance
$\varepsilon$ in a stopping rule must say whether it covers the
truncation alone, as the bound does, or all three parts. Production
implementations refine every step: the logarithm of
fdlibm~\cite{fdlibm} reduces to a narrower interval around $1$,
evaluates the same transformed series with a minimax polynomial in
place of the Taylor coefficients, and splits $\ln 2$ into a high and a
low part to control the rounding of $k \ln 2$.
Algorithm~\ref{alg:series} is the skeleton of that routine, included
so that the comparison in Section~\ref{sec:exp-series} is with a
concrete method; it is not covered by the formal verification of
Section~\ref{sec:dafny}.

A randomized method closer to the deterministic one deserves mention.
A classical Poisson generator~\cite[Chapter~X, Section~3]{Devroye86}
multiplies independent uniforms until the product falls below
$e^{-\lambda}$; the number of factors minus one is Poisson$(\lambda)$.
Supplying the threshold $1/x$ instead, for $x > 1$ with $x = 1$
returned directly, gives in the ideal model a variable with mean and
variance $\ln x$ using $\ln x + 1$ draws on average, hence an unbiased
estimator of $\ln x$ without $\gamma$ or $c(x)$; it uses real
multiplication rather than counting, and its variance $\ln x$ exceeds
$H_x - H^{(2)}_x$ by at most $1.07$ (Lemma~\ref{lem:variance-bounds}).
Logarithmic-cost randomized estimation of $\ln x$ is thus not special
to records.

\section{Selected Experiments and Reproducibility}\label{sec:experiments}

All experiments are in \texttt{experiments.py}, which uses only the
Python standard library (Python~3.14, macOS, Intel), and all raw
outputs, per replication and summarized, are in \texttt{results/};
both are in the \texttt{src} directory of the paper's public
repository.\footnote{\url{https://github.com/alidasdan/randomized-harmonic-logarithms}}
Each replication of each configuration uses its own random stream,
seeded from a string naming the experiment, the argument, and the
replication index; coverage experiments use $500$ replications and
repetition sweeps $200$. Where an experiment needs only the
distribution of $R_x$ it uses Algorithm~\ref{alg:sampler}, which has
the same distribution as the scan (Proposition~\ref{prop:sampler}).
Timings are wall-clock measurements in CPython, single runs except
where batches are stated, and are indicative only. The experiments
answer three questions; Appendix~\ref{app:tables} holds the full
tables.

\subsection{Does the theory predict the behaviour?}\label{sec:exp-theory}

Figure~\ref{fig:e6a} compares the original estimator $\Lhat_x$ (counts
from Algorithm~\ref{alg:sampler}, $n = 1000$) with the range-reduced
$\Lred_x$ (Algorithm~\ref{alg:reduced}, $q = 1$, the same $n$) on the
family $x = 3 \cdot 2^{k-1}$, $k = 2, \dots, 16$, for which $x/b =
3/2$ at every $k$: $\ln x$ grows from $1.8$ to $11.5$ while the reduced
quantity $\ln(x/b) = 0.405$ is fixed. Both standard errors follow
their theoretical values, the original growing with $x$ and the
reduced one staying below its bound. Both estimators have coverage
between $0.93$ and $0.97$ for the nominal 95\% interval, and their
root mean squared errors equal their standard errors to three
decimals, the biases being far smaller
(\texttt{results/e6a\_summary.dat}). The argument sweep of the 2017
version's figure, $x = 4^k$, gives the same agreement for $\Lhat_x$
alone (Table~\ref{tab:e1}).

\begin{figure}[htbp]
\centering
\begin{tikzpicture}
\begin{axis}[
  width=0.62\textwidth, height=0.42\textwidth,
  xmode=log, log basis x=2,
  xtick={6,24,96,384,1536,6144,24576,98304},
  xticklabels={2,4,6,8,10,12,14,16},
  xlabel={$k$, where $x = 3 \cdot 2^{k-1}$}, ylabel={standard error, $n = 1000$},
  ytick={0,0.05,0.10,0.15},
  yticklabel style={/pgf/number format/fixed, /pgf/number format/precision=2,
    /pgf/number format/fixed zerofill},
  legend pos=north west, legend cell align=left, grid=major, ymin=0, ymax=0.18]
\addplot[mark=*, blue] table[x=x, y=orig_se_emp] {results/e6a_summary.dat};
\addlegendentry{original, empirical}
\addplot[mark=none, blue, thick] table[x=x, y=orig_se_theory] {results/e6a_summary.dat};
\addlegendentry{original, theory}
\addplot[mark=square*, red] table[x=x, y=red_se_emp] {results/e6a_summary.dat};
\addlegendentry{reduced, empirical}
\addplot[mark=none, red, thick] table[x=x, y=red_se_theory] {results/e6a_summary.dat};
\addlegendentry{reduced, theory}
\addplot[mark=none, black, dashed] table[x=x, y=se_bound] {results/e6a_summary.dat};
\addlegendentry{bound}
\end{axis}
\end{tikzpicture}
\caption{Standard error of the estimate of $\ln x$ against $x = 3
  \cdot 2^{k-1}$, $k = 2, \dots, 16$, for $n = 1000$ repetitions.
  Original: $\Lhat_x$ from Algorithm~\ref{alg:sampler} counts.
  Reduced: $\Lred_x$ from Algorithm~\ref{alg:reduced} with $q = 1$.
  Empirical: the standard deviation of the estimate over $500$
  independent replications at each $x$. Theory: $\sqrt{(H_x -
  H^{(2)}_x)/n}$ for the original and $\sqrt{\Var T_{x,b}/n}$ for the
  reduced estimator. Bound: $\sqrt{\ln 2/n} = 0.0263$ (dashed).
  Empirical and theory agree within $6\%$ for the original and $7\%$
  for the reduced estimator; the relative noise of a $500$-replication
  estimate is approximately $3.2\%$, so the largest deviations are
  about two such standard errors over $15$ points. The original grows
  like $\sqrt{\ln x}$, from $0.031$ to $0.102$; the reduced estimator
  stays between $0.018$ and $0.022$, below the bound, because $x/b =
  3/2$ at every $k$.}
\label{fig:e6a}
\end{figure}

Figure~\ref{fig:sweep} shows the $n^{-1/2}$ law for both estimators:
the root mean squared error over $200$ replications against $n$ from
$10$ to $10^5$, for $\Hbar$ at $x = 256$ and for $\Lred_x$ at $x =
384$, each with its theoretical standard error. The two panels show
the same exponent at two levels; they are not an efficiency comparison
at a common argument.

\begin{figure}[htbp]
\centering
\begin{tikzpicture}
\begin{loglogaxis}[
  width=0.47\textwidth, height=0.38\textwidth,
  xlabel={repetitions $n$}, ylabel={RMSE},
  title={$\Hbar$, $x = 256$}, title style={font=\small},
  legend pos=north east, legend cell align=left, grid=major,
  xmin=7, xmax=150000]
\addplot[mark=*, blue] table[x=n, y=rmse] {results/e2_summary.dat};
\addlegendentry{empirical}
\addplot[mark=none, red, thick] table[x=n, y=se_theory] {results/e2_summary.dat};
\addlegendentry{theory}
\end{loglogaxis}
\end{tikzpicture}\hspace{0.01\textwidth}%
\begin{tikzpicture}
\begin{loglogaxis}[
  width=0.47\textwidth, height=0.38\textwidth,
  xlabel={repetitions $n$}, ylabel={RMSE},
  title={$\Lred_x$, $x = 384$}, title style={font=\small},
  legend pos=north east, legend cell align=left, grid=major,
  xmin=7, xmax=150000]
\addplot[mark=*, blue] table[x=n, y=rmse] {results/e6c_summary.dat};
\addlegendentry{empirical}
\addplot[mark=none, red, thick] table[x=n, y=se_theory] {results/e6c_summary.dat};
\addlegendentry{theory}
\addplot[mark=none, black, dashed] table[x=n, y=se_bound] {results/e6c_summary.dat};
\addlegendentry{bound}
\end{loglogaxis}
\end{tikzpicture}
\caption{Root mean squared error against the number of repetitions
  $n$, log-log axes, $200$ independent replications at each $n$ from
  $10$ to $10^5$. Left: $\Hbar$ at $x = 256$; theory is $\sqrt{(H_x -
  H^{(2)}_x)/n}$, and empirical and theory agree within $7.1\%$ at
  every $n$. Right: the range-reduced $\Lred_x$ at $x = 384$ ($b =
  256$, $q = 1$); theory is $\sqrt{\Var T_{384,256}/n} \approx
  \sqrt{0.4035/n}$, bound is $\sqrt{\ln 2/n}$ (dashed), and empirical
  and theory agree within $12\%$, the largest deviation being at $n =
  300$. The relative noise of a $200$-replication estimate of a
  standard deviation is approximately $5\%$. Both lines have slope
  $-1/2$.}
\label{fig:sweep}
\end{figure}

Table~\ref{tab:cov} shows the coverage of the nominal 95\% normal
interval in an ordinary and in a rare-tail configuration for each of
the coupled estimator of $\ln(a/b)$ and the reduced estimator. In the
ordinary cases the empirical standard error matches the theory and
coverage is about $0.94$ (with $500$ replications, the Monte Carlo
standard error of these ordinary-case coverage estimates is about one
percentage point). In the rare-tail cases, $(a, b) = (1001,
1000)$ and $x = 2^{10} + 1$, the target $\ln(a/b) \approx 0.001$ is of
the order of the standard error, in $35$ to $38\%$ of the replications
every count is zero and the interval has zero width, and coverage
falls to $0.62$--$0.65$, as Section~\ref{sec:rational} predicts; the
estimates themselves remain accurate on average. Powers of two take
the exact branch of Algorithm~\ref{alg:reduced}: under exact $\ln 2$
the output is $\ln x$, and with the stored constant it is $k \ell_2$,
which agreed with the library value at the arguments tested but is
not thereby a real-arithmetic guarantee.

\begin{table}[htbp]
\centering
\small
\begin{tabular}{llrrrrr}
\toprule
estimator & argument & target & SE emp. & SE theory & cov. & zero-width \\
\midrule
coupled $\bar T - c(a) + c(b)$ & $(a, b) = (100, 50)$ & $0.6931$ & $0.0263$ & $0.0260$ & $0.936$ & $0.000$ \\
coupled & $(a, b) = (1001, 1000)$ & $0.0010$ & $0.00096$ & $0.00100$ & $0.646$ & $0.354$ \\
reduced $\Lred_x$ & $x = 2047$ & $0.6927$ & $0.0274$ & $0.0263$ & $0.940$ & $0.000$ \\
reduced $\Lred_x$ & $x = 1025$ & $0.0010$ & $0.00097$ & $0.00099$ & $0.624$ & $0.376$ \\
\bottomrule
\end{tabular}
\caption{Coverage of the nominal 95\% normal interval, $n = 1000$,
  $500$ replications, $q = 1$ for the reduced rows. Target is the
  estimated quantity, $\ln(a/b)$ for the coupled estimator and
  $\ln(x/b)$, the sampled part of $\ln x$, for the reduced one; SE
  emp.\ is the standard deviation of the estimate across
  replications; SE theory is $\sqrt{\Var T/n}$ for the relevant tail
  count; cov.\ is the fraction of replications whose interval covers
  the true logarithm; zero-width is the fraction with zero sample
  variance. Rows are from Tables~\ref{tab:e4} and~\ref{tab:e6b}.}
\label{tab:cov}
\end{table}

\subsection{Do the improvements work?}

Table~\ref{tab:cost} compares the scan (Algorithm~\ref{alg:scan}) with
the exact sampler (Algorithm~\ref{alg:sampler}): the sampler's draws
per repetition equal its output and average $H_x$, against $x$ real
draws for the scan, and the observed means agree with $H_x$ within
the standard error of a thousand counts
(\texttt{results/e3\_summary.dat}). For the range-reduced estimator,
the recorded runs (\texttt{results/e6a\_summary.dat}) give $1.37$ to
$1.41$
integer draws per repetition on the family $x = 3 \cdot 2^{k-1}$,
against $2.4$ to $12.1$ for the original, and the boundary cases of
Table~\ref{tab:e6b} confirm zero draws at powers of two. The effect of
the scaling parameter $q$, evaluated from the finite-sum formulas, is
in Table~\ref{tab:e6d}: the bias falls like $q^{-4}$ while variance
and draws rise towards their limits. The antithetic variant of
Appendix~\ref{app:antithetic} asymptotically halves the variance per
scan (its variance is $(H_x + 1 - 2H^{(2)}_x)/2$, with ratio to
$\Var R_x$ tending to $1/2$) at the cost of a second comparison per
element, with the measured efficiency
ratios in Table~\ref{tab:e5}.

\begin{table}[htbp]
\centering
\small
\begin{tabular}{rrrr}
\toprule
$x$ & $H_x$ & scan draws per repetition & sampler draws per repetition (observed) \\
\midrule
$16$ & $3.381$ & $16$ & $3.399$ \\
$1024$ & $7.509$ & $1024$ & $7.516$ \\
$65536$ & $11.668$ & $65536$ & $11.579$ \\
\bottomrule
\end{tabular}
\caption{Cost of one repetition: the scan draws $x$ real numbers; the
  exact sampler draws one integer per record, so its draws equal its
  output, with expectation $H_x$ (Proposition~\ref{prop:sampler}).
  Observed is the mean over $n = 1000$ repetitions on one random
  stream; the standard error of that mean is between $0.04$ and  $0.10$ here. Three of the eight arguments in
  \texttt{results/e3\_summary.dat}.}
\label{tab:cost}
\end{table}

\subsection{How does the deterministic baseline compare?}\label{sec:exp-series}

Figure~\ref{fig:e7a} shows the binary64 error of Algorithm~\ref{alg:series}
against the number of terms for three arguments, measured against
$\ln x$ computed to $400$ significant digits with Python's
\texttt{decimal} module, with the uniform truncation bound. The error
falls geometrically until it reaches a rounding plateau that grows
with $k$, in line with the stored-constant term and the arithmetic
rounding of Remark~\ref{rem:ln2}, and is between $10^{-17}$ and $4
\times 10^{-15}$ across all arguments tested; the ideal truncation
error is within its bound~\eqref{eq:series-bound} in each of the $208$
configurations (Table~\ref{tab:e7a}). Compare the $n^{-1/2}$ decrease
of Figure~\ref{fig:sweep}.

\begin{figure}[htbp]
\centering
\begin{tikzpicture}
\begin{semilogyaxis}[
  width=0.62\textwidth, height=0.42\textwidth,
  xlabel={terms $N$}, ylabel={absolute error},
  legend pos=north east, legend cell align=left, grid=major,
  xmin=0, xmax=17, unbounded coords=discard]
\addplot[mark=*, blue] table[x=N, y=actual_err, discard if not={x}{2047}] {results/e7a_summary.dat};
\addlegendentry{$x = 2^{11} - 1$}
\addplot[mark=square*, red] table[x=N, y=actual_err, discard if not={x}{384}] {results/e7a_summary.dat};
\addlegendentry{$x = 384$}
\addplot[mark=triangle*, green!60!black] table[x=N, y=actual_err, discard if not={x}{1025}] {results/e7a_summary.dat};
\addlegendentry{$x = 2^{10} + 1$}
\addplot[mark=none, black, dashed] table[x=N, y=uniform_bound, discard if not={x}{384}] {results/e7a_summary.dat};
\addlegendentry{bound}
\end{semilogyaxis}
\end{tikzpicture}
\caption{Actual binary64 error of the deterministic series against the
  number of terms $N$ for $x = 2^{11} - 1$ ($z \approx 1/3$, the worst
  case), $x = 384$ ($z = 0.2$), and $x = 2^{10} + 1$ ($z \approx 5
  \times 10^{-4}$), against a $400$-digit reference, with the uniform
  truncation bound $9/(4(2N+1)3^{2N+1})$ (dashed). Each curve falls
  geometrically until it reaches a rounding plateau, here between
  $4 \times 10^{-17}$ and $4 \times 10^{-16}$, which the bound does not
  describe.}
\label{fig:e7a}
\end{figure}

Table~\ref{tab:e7b} times the series with sixteen terms, the library
logarithm, and the two randomized estimators per estimate at two
arguments, with the accuracy each attains. The series costs about
$2.5$ microseconds per estimate in CPython, twenty times the library
call, and reaches errors of $4 \times 10^{-17}$ and $1 \times
10^{-15}$; one randomized estimate with $n = 1000$ costs $0.9$ to
$1.0$ milliseconds for the reduced estimator and $2.7$ to $4.8$
milliseconds for the original, with root mean squared errors of
$0.02$ and $0.07$ to $0.10$. The timings are wall-clock medians in one
interpreter with the spread shown and are indicative; the error
columns are the measured quantities, and equal loop counts do not
imply equal accuracy.

\begin{table}[htbp]
\centering
\small
\begin{tabular}{rllrrll}
\toprule
$x$ & method & parameters & median $\mu$s & range $\mu$s & error & error measure \\
\midrule
$384$ & series & $N = 16$ & $2.48$ & $2.45$--$2.87$ & $3.9 \times 10^{-17}$ & abs.\ error \\
$384$ & \texttt{math.log} & & $0.125$ & $0.124$--$0.134$ & $3.9 \times 10^{-17}$ & abs.\ error \\
$384$ & original & $n = 1000$ & $2700$ & $2570$--$3100$ & $7.2 \times 10^{-2}$ & RMSE \\
$384$ & reduced & $n = 1000$, $q = 1$ & $915$ & $904$--$1020$ & $2.1 \times 10^{-2}$ & RMSE \\
$98304$ & series & $N = 16$ & $2.57$ & $2.51$--$2.87$ & $1.0 \times 10^{-15}$ & abs.\ error \\
$98304$ & \texttt{math.log} & & $0.124$ & $0.124$--$0.125$ & $7.4 \times 10^{-16}$ & abs.\ error \\
$98304$ & original & $n = 1000$ & $4840$ & $4770$--$5210$ & $1.0 \times 10^{-1}$ & RMSE \\
$98304$ & reduced & $n = 1000$, $q = 1$ & $964$ & $938$--$1080$ & $2.0 \times 10^{-2}$ & RMSE \\
\bottomrule
\end{tabular}
\caption{Time and accuracy per estimate of $\ln x$. Series is
  Algorithm~\ref{alg:series} with $N$ terms; \texttt{math.log} is
  the Python library call; original is $\Lhat_x$ from $n$ counts of
  Algorithm~\ref{alg:sampler}; reduced is $\Lred_x$ from
  Algorithm~\ref{alg:reduced}. Median and range are over $7$ batches
  of $20000$ calls (deterministic) or $20$ estimates (randomized),
  CPython 3.14 on an Intel Mac, including interpreter and
  \texttt{random} overhead. Error is the absolute error against the
  high-precision reference for the deterministic methods and the root  mean squared error over $500$ replications at the same $x$ and $n$
  (\texttt{results/e6a\_summary.dat}) for the randomized ones. Two of
  the three arguments in \texttt{results/e7b\_summary.dat}, where the
  $N = 8$ series is also recorded.}
\label{tab:e7b}
\end{table}

\FloatBarrier
\subsection{Formal verification and reproducibility}\label{sec:dafny}

Algorithms~\ref{alg:max}--\ref{alg:reduced} and the antithetic scan of
Appendix~\ref{app:antithetic} have been verified with the Dafny
program verifier~\cite{Leino10} (version 4.11): the deterministic
contracts of the pseudocode (outputs, draw counts, the formulas
computed) and the finite-sum identities behind the error bars (the
output distributions of Algorithms~\ref{alg:sampler}
and~\ref{alg:tail}, and the means and variances of
Propositions~\ref{prop:records}, \ref{prop:antithetic},
and~\ref{prop:reduced} and of $\Hbar$). Not verified are the modelling
layer (Proposition~\ref{prop:records} itself and the reading of a
uniform draw as a uniform mixture), the analytic Lemma~\ref{lem:bias}
and the bounds derived from it, floating point, the unbiasedness of
the sample variance, and Algorithm~\ref{alg:series}; the supplied
constant $\ell_2$ is a parameter that the proofs nowhere identify
with $\ln 2$. Appendix~\ref{app:dafny} lists the scope and assumptions
in full. The Dafny sources, \texttt{experiments.py}, the raw results
with their run records, and Python implementations of the algorithms
are in the paper's public repository (Section~\ref{sec:experiments}).

\FloatBarrier
\section{Conclusions}\label{sec:conclusions}

The practical finding is plain. A randomized estimate of $\ln x$ with
$n = 1000$ repetitions costs about a millisecond in our implementation
and has root mean squared error about $0.02$ after range reduction,
and, while sampling error dominates the bias and rounding floors, each
additional decimal digit costs a hundredfold increase in $n$;
sixteen terms of a classical series after the same reduction cost a
few microseconds and reach the rounding floor. For computing
logarithms, deterministic routines remain the right tool.

What the note offers is a self-contained example, suitable for a first course in algorithms or probability, in
which the analysis of an algorithm (the expected number of maximum
updates) is turned into an estimator of the quantity the analysis
produces; in which the same analysis supplies the exact variance and
hence honest error bars, with the sampling error and the deterministic
bias kept apart; in which one more step of the analysis (the position
of the maximum is uniform) recovers a classical exact sampler that
reduces the expected work from linear to logarithmic in $x$; and in
which coupling and a binary reduction make the sampling error bound
uniform in $x$. The probabilistic facts used are classical and
documented in the cited literature; the contribution is their
assembly, the corrections to the 2017 version, a reproducible set of
experiments that confirm the theory, and machine-checked proofs of the
algorithms' contracts and of the finite-sum identities behind the
error bars.

\section*{Acknowledgements}

The revision extends the analysis of the 2017 version. It was
prepared with two AI assistants (Anthropic's Claude and OpenAI's
Codex) working under the author's direction: the author set the
questions and the scope, and the assistants, each checking the other,
drafted and cross-checked the derivations, the exact sampler, the
experiments, and the Dafny proofs; a third assistant (Google's Gemini)
reviewed a draft. The author reviewed and decided every change, and
any remaining errors are the author's.

\appendix
\FloatBarrier

\section{Proofs and Numerical Details}\label{app:proofs}

\subsection*{Proof of Lemma~\ref{lem:bias}}

$H_x = \gamma + \psi(x+1) = \gamma + \psi(x) + 1/x$, where $\psi$ is
the digamma function. The asymptotic expansion
$\psi(x) \sim \ln x - \frac{1}{2x} - \sum_{k \geq 1}
\frac{B_{2k}}{2k\, x^{2k}}$, where the $B_{2k}$ are the Bernoulli
numbers, has the property that, for real $x > 0$,
truncating the sum leaves a remainder bounded in magnitude by the
first neglected term and of the same sign~\cite[Eq.~5.11.2 and the
remark following it]{DLMF}. With $B_2 = 1/6$ and $B_4 = -1/30$,
truncating after $k = 1$ gives $\psi(x) = \ln x - \frac{1}{2x} -
\frac{1}{12x^2} + r_x$ with $0 < r_x < \frac{1}{120x^4}$. Adding
$\gamma + 1/x$ gives~\eqref{eq:expansion}.
For monotonicity, the exact identity
\[
  r_x = \psi(x) - \ln x + \frac{1}{2x} + \frac{1}{12x^2}
\]
and the recurrence $\psi(x+1) - \psi(x) = 1/x$ eliminate the digamma
terms when consecutive remainders are subtracted:
$r_x - r_{x+1} = \ln\frac{x+1}{x} - \frac{1}{x} + \frac{1}{2x} -
\frac{1}{2(x+1)} + \frac{1}{12x^2} - \frac{1}{12(x+1)^2}$. Setting
$t = 1/x$ gives
\[
\begin{aligned}
 r_x-r_{x+1}
 &=\ln(1+t)-\frac{t}{2(1+t)}-\frac{t}{2}
   +\frac{t^2}{12}-\frac{t^2}{12(1+t)^2}\\
 &=\int_0^t\frac{u^4}{6(1+u)^3}\,du>0.
\end{aligned}
\]
The integral identity follows by differentiating the preceding
expression in $t$ and noting that both sides vanish at $t=0$. The
remainder $r_x$ approaches its bound from below as $x$ grows: the next
term of the expansion gives $r_x / (1/(120x^4)) = 1 - 10/(21x^2) +
O(x^{-4})$, which is $0.97$ at $x = 4$, so the bound is
asymptotically tight. \qed

\subsection*{Stored constants}

The constants are stored as doubles. We use the 21-digit decimal
values
\[
  \gamma \approx 0.577215664901532860606,
  \qquad
  \ln 2 \approx 0.693147180559945309417,
\]
the precision at which C library headers list such constants (the C header \texttt{math.h} defines $\ln 2$ as
\texttt{M\_LN2} but no Euler constant; the $\gamma$ value is the GNU
Scientific Library's \texttt{M\_EULER}). The decimal literals are
accurate to $5 \times 10^{-22}$; what matters is their rounding to
double precision, within half a unit in the last place ($5.6 \times
10^{-17}$ for numbers in $[1/2, 1)$). Writing $g$ for the stored
$\gamma$ and $e_\gamma = g - \gamma \approx 4.9 \times 10^{-18}$,
replacing $\gamma$ by $g$ in Algorithm~\ref{alg:scan} shifts the bias
of $\Lhat_x$ from $r_x$ to $r_x - e_\gamma$, before any other
arithmetic rounding; the magnitude of $e_\gamma$ exceeds $r_x$ beyond
approximately $x = 6400$, so for large $x$ the stored constant, not
the expansion, sets the bias in this model. The five-digit value
$0.57721$ used in the 2017 version adds $5.7 \times 10^{-6}$, which
dominates $r_x$ already for $x \geq 7$. For the stored $\ln 2$,
$\ell_2 - \ln 2 \approx -2.3 \times 10^{-17}$, so the error $k(\ell_2
- \ln 2)$ of Algorithms~\ref{alg:reduced} and~\ref{alg:series} is
about $-2.3 \times 10^{-17}\,k$: $2.8 \times 10^{-16}$ at $k = 12$,
where it first exceeds the ideal bias bound $1/(120 b^4)$ of
Algorithm~\ref{alg:reduced} with $q = 1$; $1.5 \times 10^{-15}$ at $k
= 63$; and $2.4 \times 10^{-14}$ at $k = 1023$, the largest double
exponent. The final floating-point operations introduce additional
rounding on a scale of $u \ln x$ with $u = 2^{-53}$; this scale reaches
about $5 \times 10^{-15}$ near the largest 64-bit arguments and can
exceed the stored-constant contribution. It is an order of magnitude,
not a bound on the whole sequence of operations.

\subsection*{Variance times draws}

Judged by variance times expected draws per repetition, the
range-reduced estimator costs at most $\ln 2\,(1 + \ln 2) < 1.18$ for
every $x$ and every $q$, by Proposition~\ref{prop:reduced}. The
original estimator costs $(H_x - H^{(2)}_x) H_x$, which increases
with $x$ and already equals $1.37$ at $x = 4$. At $x = 2$ the reduced
estimator draws nothing, against $0.375$; at $x = 3$, where $x/b =
3/2$, it costs at most $\ln(3/2)(1 + \ln(3/2)) < 0.57$ for every $q$
($0.296$ at $q = 1$), against $0.866$. Hence the reduced estimator is
the more efficient for every $x \geq 2$ and every $q$, and for $x = 1$
both are exact. The comparison concerns logical draws in the unit-cost
model of Section~\ref{sec:sampler}, not bit complexity.

\section{The Antithetic Variant}\label{app:antithetic}

The scan of Algorithm~\ref{alg:scan} can track the running minimum at
no extra draws. Let $R^{+}_x$ and $R^{-}_x$ be the numbers of upper
and lower records of $U_1, \dots, U_x$ and $A_x = (R^{+}_x + R^{-}_x)/2$.

\begin{proposition}\label{prop:antithetic}
$\E A_x = H_x$ and
\begin{equation}\label{eq:antivar}
  \Var A_x = \frac{H_x + 1 - 2 H^{(2)}_x}{2},
  \qquad
  \Cov(R^{+}_x, R^{-}_x) = 1 - H^{(2)}_x .
\end{equation}
In particular $\Var A_x \leq (\ln x)/2$ for every $x \geq 1$, and
$\Var A_x / \Var R_x \to 1/2$ as $x \to \infty$.
\end{proposition}

\begin{proof}
The pairs $(I^{+}_i, I^{-}_i)$ are functions of the rank $\rho_i$ of
$U_i$ among the first $i$ values ($I^{+}_i = \mathbf{1}\{\rho_i = i\}$,
$I^{-}_i = \mathbf{1}\{\rho_i = 1\}$), and the $\rho_i$ are
independent, so the pairs are independent across $i$. For $i \geq 2$
the two events are mutually exclusive with probability $1/i$ each, so
$\Cov(I^{+}_i, I^{-}_i) = -1/i^2$; for $i = 1$ both indicators equal
$1$ and the covariance is $0$. Summing gives $\Cov(R^{+}_x, R^{-}_x) =
-(H^{(2)}_x - 1)$, and $\Var A_x = \bigl(2(H_x - H^{(2)}_x) +
2(1 - H^{(2)}_x)\bigr)/4$. The bound $\Var A_x \leq (\ln x)/2$
follows from $H^{(2)}_x \geq 1$ and $H_x - 1 \leq \ln x$
(Lemma~\ref{lem:variance-bounds}): $H_x + 1 - 2H^{(2)}_x \leq H_x - 1
\leq \ln x$.
\end{proof}

The antithetic scan uses the same $x$ draws and twice the comparisons.
Its efficiency should be judged by variance times time per repetition,
which Table~\ref{tab:e5} reports; two independent scans also halve the
variance at twice the draws, and averaging two independent calls to
Algorithm~\ref{alg:sampler} gives variance $(H_x - H^{(2)}_x)/2$ using
$2H_x$ integer draws on average. The antithetic estimator has variance
lower by $(H^{(2)}_x - 1)/2$ because the two counts from one scan are
negatively correlated for $x > 1$. In Table~\ref{tab:e5}, variances
are over $2000$ repetitions and the empirical antithetic-scan variance
matches Proposition~\ref{prop:antithetic}; the time ratio is the
wall-clock time per repetition of the antithetic scan divided by that
of the plain scan, and the efficiency ratio is the corresponding ratio
of (empirical variance $\times$ time), so values below $1$ favour the
antithetic scan.

\begin{table}[htbp]
\centering
\small
\begin{tabular}{lrrrrrr}
\toprule
& \multicolumn{2}{c}{scan var.} & \multicolumn{2}{c}{antithetic scan var.} & & \\
\cmidrule(lr){2-3} \cmidrule(lr){4-5}
$x$ & theory & emp. & theory & emp. & time ratio & eff.\ ratio \\
\midrule
\eFiveRows
\bottomrule
\end{tabular}
\caption{Plain scan versus antithetic scan, $2000$ repetitions each.
  Theory is the variance of one repetition, $H_x - H^{(2)}_x$ for the
  plain scan and~\eqref{eq:antivar} for the antithetic scan; emp.\ is
  the sample variance over the repetitions; time ratio is the
  wall-clock time per repetition of the antithetic scan divided by
  that of the plain scan (single measurements, indicative); eff.\
  ratio is the corresponding antithetic-to-plain ratio of empirical
  variance $\times$ time per repetition, so values below $1$ favour
  the antithetic scan.}
\label{tab:e5}
\end{table}

\FloatBarrier
\section{Additional Experimental Tables}\label{app:tables}

All tables come from the recorded runs described in
Section~\ref{sec:experiments}; raw per-replication files and run
records are in the repository.

\begin{table}[htbp]
\centering
\small
\begin{tabular}{lrrrrrrrrr}
\toprule
$x$ & $H_x$ & $\Hbar$ & SE emp. & SE theory & $\Lhat_x$ & $\ln x$ &
rel.\ err.\ \% & cov.\ $H$ & cov.\ $\ln$ \\
\midrule
\eOneRows
\bottomrule
\end{tabular}
\caption{Argument sweep, $x = 4^k$, $n = 1000$, $500$ replications.
  $\Hbar$ and $\Lhat_x$ are averaged over replications; SE emp.\ is
  the standard deviation of $\Hbar$ across replications; SE theory is
  $\sqrt{(H_x - H^{(2)}_x)/n}$; rel.\ err.\ is $|\Lhat_x - \ln
  x|/\ln x$ in percent for the seed-averaged $\Lhat_x$ (hence of $5
  \times 10^5$ counts; a single run of $n = 1000$ has relative
  standard error $1.9\%$ at $x = 4$ falling to $0.9\%$ at $x =
  65536$); cov.\ $H$ and cov.\ $\ln$ are the fractions of
  replications whose nominal 95\% interval $\Hbar \pm
  1.96\sqrt{s^2/n}$ covers $H_x$ and whose shifted interval covers
  $\ln x$. They nearly coincide because $r_x$ is far below the
  interval width. The empirical and theoretical standard errors agree
  within $4.2\%$; the relative noise of a $500$-replication estimate
  of a standard deviation is approximately $3.2\%$.}
\label{tab:e1}
\end{table}

\begin{table}[htbp]
\centering
\footnotesize
\setlength{\tabcolsep}{3.5pt}
\begin{tabular}{rrrrrrrrrrl}
\toprule
& & & \multicolumn{4}{c}{coupled (Alg.~\ref{alg:tail})} & \multicolumn{3}{c}{independent difference} & \\
\cmidrule(lr){4-7} \cmidrule(lr){8-10}
$a$ & $b$ & $\ln(a/b)$ & mean & SE emp. & SE theory & cov. & SE emp. & SE theory & cov. & bias \\
\midrule
\eFourRows
\bottomrule
\end{tabular}
\caption{Estimates of $\ln(a/b)$ with $n = 1000$, $500$ replications,
  by the coupled tail sampler and by the difference of two independent
  estimates of $H_a$ and $H_b$. Mean is the coupled estimate averaged
  over replications; SE emp.\ is the standard deviation of the
  estimate across replications; SE theory is $\sqrt{\Var T_{a,b}/n}$
  for the coupled estimator and $\sqrt{(\Var R_a + \Var R_b)/n}$ for
  the difference; cov.\ is the fraction of replications whose nominal
  95\% normal interval covers $\ln(a/b)$; bias is $r_a - r_b$ in
  50-digit arithmetic, appreciable only for $b = 1$. At $(1000, 999)$
  the coupled estimator is two orders of magnitude more precise than
  the difference; the pairs $(1000, 999)$ and $(1001, 1000)$ are the
  rare-tail regime of Section~\ref{sec:rational}, where the interval
  has zero width in $35$ to $37\%$ of replications.}
\label{tab:e4}
\end{table}

\begin{table}[htbp]
\centering
\small
\begin{tabular}{rrrrrrrrrl}
\toprule
$x$ & $k$ & $b$ & $\ln(x/b)$ & SE emp. & SE theory & draws & cov. &
zero & bias \\
\midrule
\eSixBRows
\bottomrule
\end{tabular}
\caption{Range-reduced estimator at boundary arguments, $n = 1000$,
  $500$ replications, $q = 1$. $k$ and $b = 2^k$ are the exponent and
  the reference power of two with $b \leq x < 2b$; SE emp., SE theory,  draws, and cov.\ are as in Figure~\ref{fig:e6a} and Table~\ref{tab:cov}
  (SE emp.\ and SE theory), with draws the mean number of integer draws
  per repetition; zero is the fraction
  of replications with a zero-width interval (one by construction at
  powers of two, where no counts are sampled); bias is $r_x - r_b$ in
  50-digit arithmetic, exactly $0$ at powers of two. At powers of two
  the output $k \ell_2$ with the stored $\ell_2$ coincided with the
  library value of $\ln x$ in every replication; the $9 \times
  10^{-16}$ discrepancy in the aggregate mean at $x = 1024$ in the
  summary file is rounding in the average over $500$ identical values.
  At $x = 2047$ the theoretical standard error is within $0.1\%$ of
  the bound $0.02633$. At $x = 3$ the bias $-3.7 \times 10^{-4}$ is
  consistent with the averaged error $-6.5 \times 10^{-4} \pm 6.8
  \times 10^{-4}$ over the replications.}
\label{tab:e6b}
\end{table}

\begin{table}[htbp]
\centering
\small
\begin{tabular}{rrrrllrrr}
\toprule
$x$ & $q$ & $qx$ & $qb$ & bias & bound & $\Var T_{qx,qb}$ &
$\ln(x/b)$ & draws \\
\midrule
\eSixDRows
\bottomrule
\end{tabular}
\caption{Effect of the scaling parameter $q$ on the range-reduced
  estimator, evaluated from the finite-sum formulas. Bias is $r_{qx} -
  r_{qb}$ in 50-digit arithmetic and bound is $1/(120 (qb)^4)$; the
  variance per repetition is bounded by $\ln(x/b)$ and the expected
  draws per repetition, $1 + H_{qx} - H_{qb}$, by $1 + \ln(x/b)$. The
  bias falls like $q^{-4}$ while variance and draws rise towards their
  limits.}
\label{tab:e6d}
\end{table}

\begin{table}[htbp]
\centering
\small
\setlength{\tabcolsep}{4pt}
\begin{tabular}{lrrrrrrr}
\toprule
 & & \multicolumn{5}{c}{actual error after $N$ terms} & bound \\
\cmidrule(lr){3-7}
$x$ & $z$ & $N = 1$ & $N = 2$ & $N = 4$ & $N = 8$ & $N = 16$ & $N = 16$ \\
\midrule
\eSevenARows
\bottomrule
\end{tabular}
\caption{Accuracy of the deterministic series (Algorithm~\ref{alg:series})
  in binary64 on the family $x = 3 \cdot 2^{k-1}$, the boundary
  arguments of Table~\ref{tab:e6b}, and three large integers. $z = (x
  - b)/(x + b)$; actual error is $|\mathrm{fl}(L_N) - \ln x|$ against
  a $400$-digit reference (unchanged to $395$ digits when recomputed
  at $440$), with the binary64 output converted to decimal exactly;
  bound is the ideal truncation bound $2 z^{2N+1}/((2N+1)(1 - z^2))$
  of~\eqref{eq:series-bound} at $N = 16$ (zero at powers of two). The
  ideal truncation error, computed directly as the positive series
  tail at $400$ digits with the omitted remainder bounded by $5 \times
  10^{-243}$, is within its bound in each of the $208$ configurations
  by exact decimal comparison. The errors at $N = 16$ are the rounding
  plateau, not truncation; at powers of two the series contributes
  nothing and the error is that of $k$ times the stored $\ln 2$. Full
  data for $N = 1, \dots, 16$ are in \texttt{results/e7a\_summary.dat}.}
\label{tab:e7a}
\end{table}

\FloatBarrier
\section{Scope of the Formal Verification}\label{app:dafny}

The Dafny development (version 4.11; \texttt{dafny/} in the
repository, with a README and a recorded run of the full suite)
verifies the following. On the deterministic side:
Algorithm~\ref{alg:max} returns the maximum and executes its update
line exactly $R_x$ times; Algorithm~\ref{alg:scan} makes $nx$ draws,
each count equals the record count of the drawn list, and the mean,
the sample variance, and the correction are computed as displayed,
with $\ln 1 = 0$ returned exactly; Algorithms~\ref{alg:sampler}
and~\ref{alg:tail} make exactly the numbers of draws claimed in
Propositions~\ref{prop:sampler} and~\ref{prop:tail}, and compute $R_x$
and $T_{a,b}$ exactly when the uniform draw is replaced by the position
of the maximum; Algorithm~\ref{alg:reduced} computes $k$ and $b = 2^k$
with $b \leq x < 2b$ (the pair is unique), returns $k \ell_2$ with
zero reported variance and no draws when $x$ is a power of two, and
otherwise makes exactly $n$ calls of Algorithm~\ref{alg:tail} on $(qx,
qb)$, returns $k \ell_2 + \bar T - c(qx) + c(qb)$ with the
Bessel-corrected $s^2/n$, and uses exactly $n$ plus the sum of the
counts integer draws, where the supplied constant $\ell_2$ is a
parameter that the proofs nowhere identify with $\ln 2$. On the
distributional side: the output distributions of
Algorithms~\ref{alg:sampler} and~\ref{alg:tail}, written as recurrences
that follow the loops branch by branch, equal the distributions of
$R_x$ and $T_{a,b}$; the mean and variance in
Proposition~\ref{prop:records} and~\eqref{eq:tail}, the covariance and
variance in Proposition~\ref{prop:antithetic}, the sign of the gap
$2\Var R_b$ in Section~\ref{sec:rational}, the mean $H_x$ and variance
$(H_x - H^{(2)}_x)/n$ of $\Hbar$ over $n$ repetitions, and for
Algorithm~\ref{alg:reduced} with $x > b$ the mean $k \ell_2 + H_{qx} -
H_{qb} - c(qx) + c(qb)$ and centered second moment $\Var T_{qx,qb}/n$
of the estimate over $n$ repetitions, with the additive shift carried
through the sums, together with the expected draw count $n(1 + H_{qx}
- H_{qb})$, are proved as identities between finite sums.

What the verifier does not check is the modelling layer:
Proposition~\ref{prop:records} itself (the independence of the record
indicators, which comes from the exchangeability of continuous draws),
the reading of a uniform draw as a uniform mixture when the output
distribution of a loop is written down, an axiomatized square root,
and the analytic Lemma~\ref{lem:bias}. Floating point, the
unbiasedness of the sample variance, the elementary variance bounds of
Lemma~\ref{lem:variance-bounds} and their consequences, the
asymptotic ratio in Proposition~\ref{prop:antithetic}, the sign and
bounds of the bias of Algorithm~\ref{alg:reduced}, the uniform bounds
of Section~\ref{sec:reduction}, and the Bernstein rules are not
formalized; one conditional lemma records that, with the identities
\eqref{eq:reduce} and~\eqref{eq:lnab} taken as hypotheses, the mean of
$\Lred_x$ minus $\ln x$ equals $r_{qx} - r_{qb}$.
Algorithm~\ref{alg:series} and Proposition~\ref{prop:series} are not
covered by the proofs. The only termination assumption concerns the
demonstration random source, which has no loops.

\section{Revision Note}\label{app:revision}

The 2017 version of this note~\cite{Dasdan17} described the scan
estimator with the correction $1/(2x)$ and a five-digit value of
$\gamma$, and reported a few numerical results. This revision keeps
that core, improves the correction and the constant
(Section~\ref{sec:accuracy} and Appendix~\ref{app:proofs}), uses the
Bessel-corrected sample variance, and adds everything else. Two
statements in the 2017 version are corrected.
\begin{itemize}
\item Its abstract said that $H_x - \ln x$ tends to zero; it tends to
  $\gamma$.
\item It described the algorithm as parallelizable by splitting the
  list and merging the pieces, because the maximum of a list is the
  maximum of the maxima of its parts. The maximum does combine, but
  the record count does not: the blocks $[3,1]$ and $[2,4]$ have one
  and two records, while their concatenation $[3,1,2,4]$ has two, not
  three. (A parallel prefix-maximum computation can recover the
  records, but that is a different algorithm.) The outer loop over
  repetitions parallelizes directly.
\end{itemize}
One statement invited misreading. The 2017 version said that the
algorithm uses integer arithmetic only, apart from the final
averaging; this refers to the counting, and the comparisons in the
scan are between real numbers. The integer sampler of
Section~\ref{sec:sampler} removes even those: it draws and compares
integers only. Two remarks of an earlier draft of this revision, on
estimating $\log_2 x$ and on reducing to a finer interval with a
second supplied constant, were omitted as asides and are recoverable
from the repository's history.

\end{document}